\documentclass[a4paper]{article}
\usepackage[utf8]{inputenc}
\usepackage[T1]{fontenc}
\usepackage{lmodern}
\usepackage[ruled,vlined]{algorithm2e}
\usepackage{graphicx}
\usepackage{amsfonts}
\usepackage{amsmath}
\usepackage{xcolor}
\usepackage{url}
\usepackage{subcaption}
\usepackage{color,soul}
\usepackage{newtxtext,newtxmath}
\usepackage[colorlinks=true,citecolor=red,linkcolor=black, bookmarks=false,naturalnames]{hyperref}
\usepackage[UKenglish]{babel}
\usepackage{eurosym}

\usepackage{geometry}
\newcommand{\R}{\mathbb{R}}

\newcommand{\vect}[1]{\mathbf{#1}}

\begin{document}
\title{Efficient Numerical Calibration of Water Delivery Network Using Short-Burst Hydrant Trials}

\author{Katarzyna Kołodziej, Michał Cholewa, Przemysław Głomb, \\ Wojciech Koral, Michał Romaszewski}

\author{
  Katarzyna Kołodziej\textsuperscript{1}\textsuperscript{*}, Michał Cholewa\textsuperscript{1}, Przemysław Głomb\textsuperscript{1},\\ Wojciech Koral\textsuperscript{2}, Michał Romaszewski\textsuperscript{1}\\ 
  \textsuperscript{1} Institute of Theoretical and Applied Informatics,\\ Polish Academy of Sciences,\\ Bałtycka 5, 44-100 Gliwice, Poland\\
  \textsuperscript{2} AIUT Sp. z o.o., Wyczółkowskiego 113,  44-109, Gliwice, Poland\\
\textsuperscript{*}Corresponding author: \texttt{kkolodziej@iitis.pl} 
}
\date{\today}

\maketitle

\begin{abstract}
    Calibration is a critical process for reducing uncertainty in Water Distribution Network Hydraulic Models (WDN HM). However, features of certain WDNs, such as oversized pipelines, lead to shallow pressure gradients under normal daily conditions, posing a challenge for effective calibration. This study proposes a calibration methodology using short hydrant trials conducted at night, which increase the pressure gradient in the WDN. The data is resampled to align with hourly consumption patterns. In a unique real-world case study of a WDN zone, we demonstrate the statistically significant superiority of our method compared to calibration based on daily usage. The experimental methodology, inspired by a machine learning cross-validation framework, utilises two state-of-the-art calibration algorithms, achieving a reduction in absolute error of up to 45\% in the best scenario.
\end{abstract}

\section{Introduction}\label{sec:introduction}
Hydraulic models are essential tools for diagnosing issues in Water Distribution Networks (WDNs), such as leak detection \cite{hu2021novel}, leak localisation \cite{romero2022leak}, and modelling incompressible unsteady flow \cite{kocsucu2022extending}. These models must produce reliable simulation results. Consequently, calibration is required during the development and validation of hydraulic models \cite{walski2017calibration}. Calibration is crucial for adapting the model to actual network operating conditions, thereby reducing uncertainty.

The calibration process involves reconciling measurements, typically point pressure or pipe flow rates, with simulation results by adjusting uncertain hydraulic model parameters. Key parameters include junction demands, control settings at regulating devices, and pipe roughness coefficients \cite{savic2009quo}, which have a significant impact on head loss \cite{abdulameer2022comparison}. Roughness is rarely determined empirically \cite{santos2020simultaneous} as it depends on various factors such as pipe material, diameter, age, curvature, and the degree of erosion \cite{malek2020factors}.
Currently, most calibration algorithms are optimisers which can be divided into non-evolutionary (gradient) ones -- appreciated for their simplicity but exposed to locally optimal solutions \cite{zhao2022simplerisbetter} -- and evolutionary ones, able to find (near-) optimal solutions in complex problems \cite{savic2009quo}. Among non-evolutionary algorithms, we can distinguish Gauss-Newton algorithm \cite{piller2017sensitivity}, Levenberg-Marquardt algorithm \cite{steffelbauer2022pressure} or Sequential Least Squares Programming (SLSQP) \cite{zhao2022simplerisbetter}. Evolutionary algorithms include Genetic Algorithm \cite{zanfei2020calibr}, Differential Evolution \cite{zhao2022simplerisbetter} or Particle Swarm Optimisation (PSO) \cite{moghaddam2020pso}.

The calibration process presents several challenges, one of which is the scarcity of measurement data. Due to the high cost of equipment, WDNs typically have a limited number of sensors. In urban networks, only 3-4\% of nodes are equipped with pressure sensors. For instance, in the L-Town network \cite{vrachimis2022battledim}, 33 sensors cover 782 nodes; in a Zhejiang Province city network \cite{shao2019time}, there are 20 sensors for 492 nodes, and in the Modena network \cite{zanfei2020calibration}, ten sensors monitor 267 nodes. Furthermore, the placement of these sensors is often suboptimal, providing an incomplete representation of the network's operational conditions \cite{huang2024placement}.
For metaheuristic evolutionary calibration algorithms, this presents a significant issue, leading to multiple potential solutions \cite{zhao2022simplerisbetter}. Additionally, calibration is hindered by the presence of oversized pipelines in many WDNs, a common practice to accommodate high water flow rates required for firefighting \cite{BERARDI2017feasibility}. These oversized pipes reduce head loss and water velocity \cite{abdulameer2022comparison}. In such cases, even substantial adjustments to roughness coefficients have minimal impact on head loss \cite{walski2017calibration}. These challenges necessitate tailored solutions.

Machine learning has recently been applied to address data scarcity in WDNs. In \cite{meirelles2017}, the authors proposed a calibration method combining a multi-layer perceptron artificial neural network (MLP-ANN) with Particle Swarm Optimisation (PSO). The MLP-ANN estimates network-wide pressure using limited measurements, reducing the degrees of freedom in the optimisation process. PSO then optimises using pressure data estimated for the entire network rather than a subset of nodes. Similarly, \cite{ZANFEI2023120264} introduced an algorithm combining a Graph Neural Network (GNN) with a Genetic Algorithm (GA), arguing that GNN can better capture the morphological structure of WDNs and improve pressure gradient reconstruction.

Another promising approach involves calibrating models after grouping or clustering pipes based on selected decision variables, reducing the optimisation problem's complexity. This was effectively applied in \cite{zhao2022simplerisbetter}, where different formulations of grouping decision variables were tested with three optimisers, demonstrating improved computational efficiency with clustering. Similarly, \cite{chen2022clustering} introduced Sensitivity-Oriented Clustering, which groups pipe roughness coefficients according to their impact on nodal pressure and pipe flow rates.

To address the issue of low-pressure gradients in WDNs, hydrant trials are recommended \cite{walski2017calibration}. Designed initially as fire hydrant performance tests, Polish water utilities regularly conduct hydrant trials to ensure hydrants are operational for fire protection \cite{regulation2009}. These trials involve controlled high-capacity water discharges from fire hydrants, inducing high-velocity flows and increasing head loss in a subset of WDN pipelines \cite{walski2017calibration}, which can improve calibration accuracy \cite{lippacher2019hp}. However, the lack of clear guidelines regarding the optimal length and discharge volumes for hydrant trials poses a challenge for effective calibration. Hydrant trial durations vary significantly depending on the purpose; for instance, trials for leak detection range from 5 minutes \cite{suchorab2021water} to over 17 hours \cite{machell2010online}. For calibration purposes, \cite{sophocleous2017two} reported four parallel trials lasting 4.5 hours. These surveys can be expensive, making them impractical for many water utilities.

This article introduces a calibration method that utilises a short (several-minute) impulse from nightly fire hydrant trials, resampled to match hourly consumption data. We demonstrate that this method achieves hydraulic model calibration results that are comparable to, or better than, those obtained from traditional daytime-based calibration. The method is applied to a case study of a complex Water Distribution Network (WDN) zone in a Polish city.  

Our contributions are:
\begin{enumerate} 
\item We propose a method for calibrating a District Metered Area (DMA) in a single night, using a limited number of temporarily mounted hydrant pressure sensors with a high sampling rate (1 second) during hydrant trials. 
\item We demonstrate that a short (lasting a few minutes) flow-inducing pulse, combined with hourly water consumption data, allows for more effective model calibration than a conventional method during peak daily consumption. This approach is particularly effective for hydrant trials conducted at locations far from the reservoir (at the end of the DMA). We test this method on two state-of-the-art calibration algorithms that effectively address the data scarcity problem. 
\item We present a unique case study of a challenging DMA involving four hydrant trials and experiments using a machine learning-inspired cross-validation methodology. 
\end{enumerate}

\section{Methods and materials}

In this section, we present the hydraulic model of the Water Distribution Network (WDN) zone selected for the experiment based on a unique data set not previously described. We outline the data used for calibration and the WDN's data collection procedure, as well as the two state-of-the-art calibration methods employed in our experiments.

\subsection{Study area} \label{sec:study_area}
\begin{figure}[ht]
    \centering
    \includegraphics[scale=0.7]{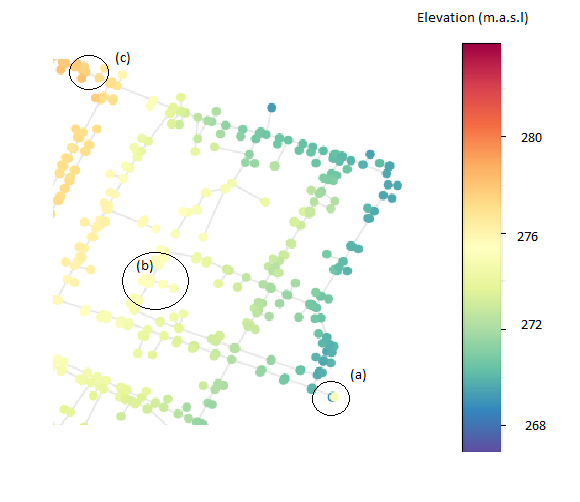}
    \caption[The entrance to the study area]{The entrance to the study area, coloured by elevation above sea level. Key regions include: (a) the reservoir, (b) an area of rapid residential expansion with a subsystem of pipes of varying diameters and materials, and (c) the main pipe connecting to the rest of the DMA to the west.}
    \label{fig:enter-label}
\end{figure}

For our study, we have selected a DMA of an industrial city in southern Poland, supplied from a single reservoir (the water tank in the inflow of the WDN zone, which stabilizes the pressure in this zone) with a hydraulic head of about $305.6\: \mathrm{mH_2O}$. The WDN zone hydraulic model consists of $n = 1071$ nodes representing junctions and $l = 1086$ links representing pipes. An inflow fragment of the DMA is visualized in Fig.~\ref{fig:enter-label}. 

The district under study is primarily residential, with around 300 active households during the midday peak. Maximum individual usage is $0.52\:\mathrm{m^3/h}$, and average peak inflow is approximately $9.05\:\mathrm{m^3/h}$ under normal conditions.

The pipes in this area have larger diameters relative to water usage compared to many reference models, such as \cite{vrachimis2022battledim} and \cite{vertommen2021robust}. For example, the study area's pipe diameter-to-demand ratio is $13.55\:\mathrm{\frac{m}{m^3/h}}$, nearly 24 times higher than the L-Town benchmark of $0.57\:\mathrm{\frac{m}{m^3/h}}$ \cite{vrachimis2022battledim}. This results in much lower average flow rates ($0.000127\:\mathrm{m^3/s}$) compared to L-Town's $0.00122\:\mathrm{m^3/s}$, creating significant challenges for calibration. 

\subsubsection{Data set} \label{sec:dataset}
The data used for calibration experiments involves pressure head, flow rate, and demands. It was collected during the two-day survey performed in the considered WDN zone. The survey registered DMA's normal operating conditions as well as a set of short-lasting night fire hydrant trials. 
Hydrant trials were conducted at night during the minimum night flow (MNF) period when water consumption in the DMA is typically at its lowest. A temporary setup designed by one of the authors was used to measure hydrant discharge capacity. This setup included a battery-powered electromagnetic flow meter (in a constricted version, not requiring straight sections before/after the flow meter), a high-precision pressure transducer, a gate or ball valve for flow adjustment, and a recorder with at least one-second resolution. Connected to the hydrant via fire hoses, this system enabled precise estimation of the hydrant's discharge capacity.
The discharge was conducted for several minutes at each hydrant location. The sizes and durations of the hydrant trials for calibration are as follows: \textbf{Trial H0}: Flow rate 9.684 $\mathrm{m^3/h}$, duration 240 s; \textbf{Trial H1}: Flow rate 9.615 $\mathrm{m^3/h}$, duration 180 s; \textbf{Trial H2}: Flow rate 10.06 $\mathrm{m^3/h}$, duration 240 s; \textbf{Trial H3}: Flow rate 10.56 $\mathrm{m^3/h}$, duration 180 s.

Hydrant trials H0 and H1 are located in the midwest and northwest parts of the WDN, respectively, far from the reservoir, and are referred to as '\textit{far}' trials. Hydrant trials H2 and H3, situated less than several hundred metres from the reservoir, are referred to as '\textit{close}' trials.

During the measurements, $m=11$ temporary pressure sensors with a 1-second sampling rate were installed in the WDN zone. Due to constraints, sensors could only be mounted on fire hydrants, limiting optimal coverage of the entire zone. Sensor placement was based on the operational staff's experience. Additionally, a second-sampling sensor was placed near the reservoir to monitor the reservoir pressure head. Precise elevation measurements were taken at the sensor installation sites.

\begin{figure}[ht]
\begin{subfigure}{0.49\textwidth}
\includegraphics[width=1\textwidth]{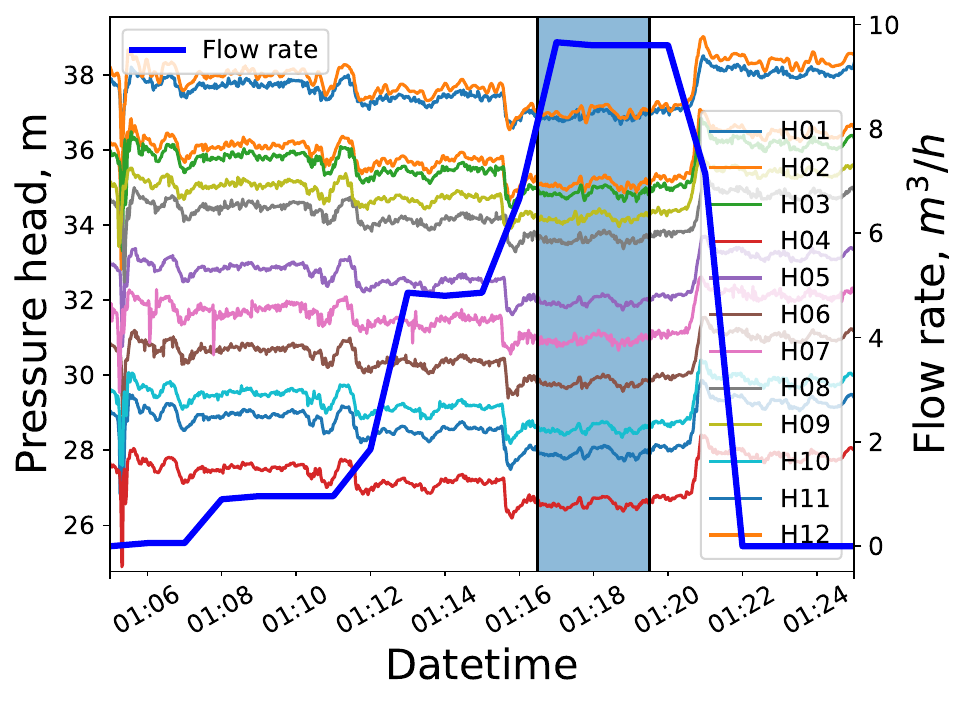}
\subcaption[]{}
\label{fig:hydrant_trial_2c}
\end{subfigure}
\begin{subfigure}{0.49\textwidth}
\includegraphics[width=1\textwidth]{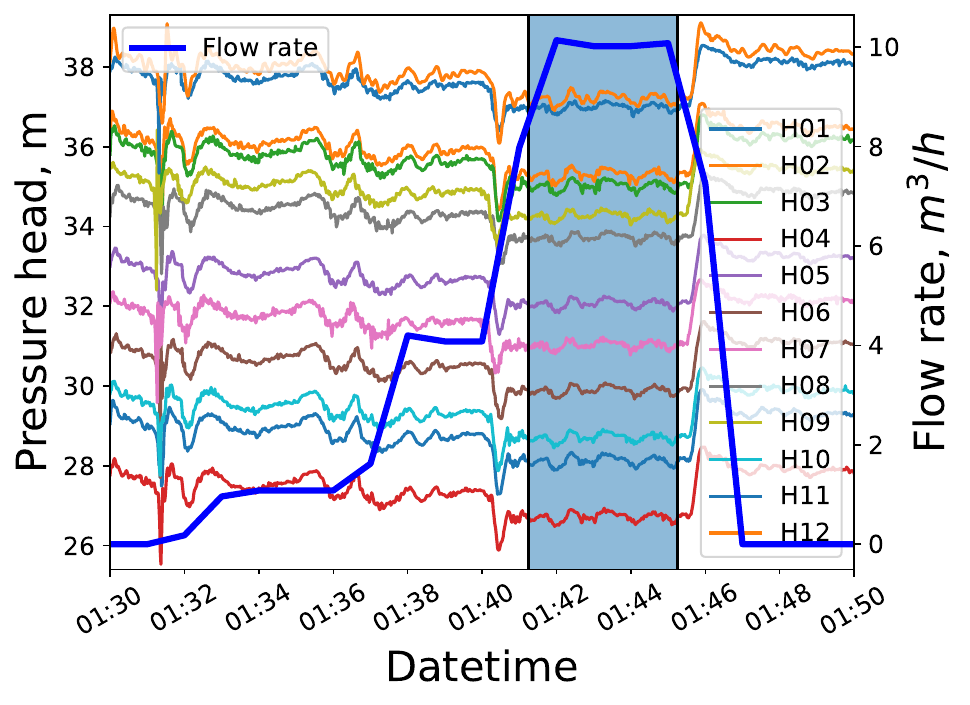}
\subcaption[]{}
\label{fig:hydrant_trial_3c}
\end{subfigure}
\caption[Pressure head and flow rate during two subsequent hydrant trials performed in the study area]{Pressure head and flow rate during two subsequent hydrant trials performed in the WDN zone. The peak water discharge during hydrant probes is marked by the blue rectangle between (a) 1:17 and 1:19, and (b) 1:41 and 1:45. In both cases, the water flow rate (blue line) in the WDN rises from nearly no usage to around $10\: \mathrm{m^3/h}$, which was the maximum flow rate set on the hydrants. Also, both trials indicate a significant drop in pressure head on all sensors.} 
\label{fig: hydrant_trials}
\end{figure}

Figure \ref{fig: hydrant_trials} presents pressure head and flow rate data visualisations for two sample hydrant trials. A few seconds of delay were observed between the stabilisation of the pressure head and flow rate data. Therefore, pressure head stabilisation was used to determine the data period for calibration.
Hourly water consumption data for the WDN zone was collected using stationary water meters. To align the 1-second pressure head data with the hourly demand data, we conduct preprocessing in the following way:
\begin{enumerate}
\item Set the time step to 1 hour, which is the lowest measurement sampling resolution;
\item Define each hydrant trial period based on the pressure head stabilisation criterion;
\item Extract pressure head readings from all sensors during the stabilised period and average each sensor's pressure head data;
\item Extrapolate the averaged pressure data to the 1-hour time step.       
\end{enumerate}

\subsection{Methods}
\subsubsection{Hydraulic simulation}
\paragraph{Hydraulic model} Let the hydraulic model be denoted as a graph $\mathcal{G} = G(N, L)$, where $N$ is a set of $n$ nodes representing junctions and $L$ is a set of $l$ links representing pipes. Let $M \subset N$ denote a subset of $m$ nodes where pressure sensors are installed, and $Q \subset N$ denote a subset of $q$ nodes with demands. We assume that node elevations and pipe diameters are constant. 

Let  $\vect{p}$ denote the vector of pressure head values for the entire set $N$. The term 'pressure head' refers to the height of the liquid column that exerts particular static pressure. Let $\vect{p}^M$ be the pressure head vector for the nodes belonging to a subset of installed sensors $M$ and $\vect{p}^Q$ be the pressure vector for the nodes belonging to the registered demand subset $Q$.

\paragraph{Simulation} The input data for the hydraulic model simulation is the hydraulic head in the reservoir $h^R$, the demand vector at the end demand nodes $\vect{d} \in \mathbb{R}^q$  and the pipe roughness coefficients vector $\vect{r} \in \mathbb{R}^l$. The simulation $\mathbb{S}$ allows for determining the pressure head vector $\vect{p}\in \mathbb{R}^n$:
\begin{equation}
    \mathbb{S}_{\mathcal{G}}(h^R, \vect{d}, \vect{r}) = \vect{p}
    \label{eq:model}
\end{equation}

Let $s_i=(h^R, \vect{d})_i, i=\{1,2,..., k\}$ be one of the $k$ scenarios in the form of model input data. For input data $s_i$, if the roughness vector $\vect{r}$ is known, it is possible to generate a simulated pressure head vector describing $i$-th scenario, denoted as $\vect{p}_i$.

As stated in Sec.~\ref{sec:dataset}, pressure head measurements from the time of hydrant trials are extrapolated to 1 hour period consistent with the simulation time step. This method is also applied to prepare the reservoir head as input data for hydraulic simulation. The reservoir head is calculated as follows:
\begin{equation}
    h^R=p^R+e^R
    \label{eq:res_head}
\end{equation}
where $p^R$ is the pressure head in the reservoir (from reservoir sensor readings), and $e^R$ is the reservoir elevation head.

The WDN zone hydraulic model was created using EPANET software \cite{rossmanepanet}. To simulate events of this model, we use the EpanetSimulator from the Water Network Tool for Resilience (WNTR) Python package \cite{klise2020water}.

\subsubsection{Calibration procedure} \label{sec:calibration_procedure}

We perform the calibration procedure given the following input data:
initial roughness vector $\vect{r}^0$, determined by an expert based on the pipes material properties;
a family of scenarios $\{s_i\}_{i=1}^k=\{(h^R, \vect{d})_i\}_{i=1}^k$;
a family of reference pressure elevation vectors $\{\vect{p}^M_{i} \}_{i=1}^k, \vect{p}^M_{i} \in \mathbb{R}^m$;
parameter $J \subseteq M$ defining the subset of reference nodes considered in calibration;
parameter $O \subseteq N$ defining the subset of $o$ nodes used for optimization stage.  We use two state-of-the-art calibration approaches.
\paragraph{ANN-PSO method}

We utilize the algorithm proposed in \cite{meirelles2017} in the following way:

\begin{enumerate}
\item ANN training stage: For each scenario $s_i$, a separate ANN model is trained. First, a large number of simulated scenarios is created based on $s_i$ in the following way: roughness coefficient values are randomly selected from the given value range, and demands are randomly disturbed in such a way that they sum up to total real-life consumption calculated from measurements (this is the exception from the original description in \cite{meirelles2017}, but it allows for realistic demands manipulation). Then, training and validation sets are built based on pressure head simulation results. The ANN is trained, and its hyperparameters are decided on the validation set;
\item ANN inference stage: For each $s_i$, preprocessing is conducted using the ANN with parameter $O \subseteq N$, aimed at expanding the set $\vect{p}^M_{J,i}$ to $\vect{p}_{O,i}$ using the scenario data and the default roughness set $\vect{r}^0$:
		\begin{equation}
			ANN: s_i,\vect{r}^0, \vect{p}^M_{J, i},  \rightarrow \vect{p}_{O, i}
		\end{equation}
\item PSO stage: the optimization procedure is performed using the family $\{\vect{p}_{O,i}\}_{i=1}^k$:
		\begin{equation}
			OPT:\{\vect{p}_{O, i}\}_{i=1}^k, \mathcal{G}, \vect{r}^0,  \{s_i\}_{i=1}^k \rightarrow \vect{r}
		\end{equation}
\end{enumerate}

The PSO stage was adapted to the needs arising from the data provided. We performed PSO parameters selection, which is described in Sec.~\ref{sec:PSO_params}. We decided to treat the optimisation problem as constrained and applied the Bratton-Kennedy boundary handling method \cite{bratton2007defining}.

\paragraph{Clustering-COBYLA method}

This algorithm, derived from~\cite{zhao2022simplerisbetter}, is based on two steps: clustering links and usage of a classical optimization algorithm on the reduced (clustered) graph. The algorithm iteratively optimises a reduced roughness vector based on selected error criterion (e.g. mean absolute error, MAE). The procedure is as follows:

\begin{enumerate}
\item Obtain reduced dimension roughness vector from initial roughness vector by performing KMeans on pipes based on error criterion (by default: both pipe roughness and flow features, standardised);
\item Calibrate based on mean MAE among all graphs (observed vs simulated pressure heads) -- minimise mean MAE among graphs by iteratively changing the reduced roughness vector and comparing the difference of observed and simulated pressure heads on sensor points.
\end{enumerate}

Initially, \cite {zhao2022simplerisbetter} used the SLSQP algorithm, but there were problems with the gradient in our data. The problem was traced to the low absolute values of the gradient. As COBYLA is a well-established algorithm which is also gradient-free, it performs as well as the original method and fits within the simplicity assumption of~\cite{zhao2022simplerisbetter}. Using roughness and flow is a natural and robust grouping criterion from~\cite{zhao2022simplerisbetter}, and it provides a natural fit for the KMeans algorithm. Features are standardized before grouping. The number of clusters was selected as a good choice after trying several criteria and obtaining similar values. As the original approach to minimization was unbounded, the implementation was adjusted to add bounds as constraints.

\section{Experiments} \label{sec:experiments}

Our experiments aim to evaluate the generalisation ability of the calibration procedure -- specifically, whether the hydraulic model calibration reduces the error between simulated and measured pressure head values when tested on previously unseen data.

We employ K-fold cross-validation \cite{berrar2019cross}, a common machine-learning evaluation technique. The data set is divided into $k$ subsets (folds); the model is trained on $k-1$ folds and tested on the remaining fold. This process is repeated for each fold.

The main experiment (Sec.~\ref{sec:main_exp}), referred to as leave-one-situation-out, involves calibrating the model with a set of scenarios and testing it on the scenario left out.

Additionally, we perform a supplementary experiment (Sec.~\ref{sec:suppl_exp}), called leave-one-sensor-out, to ensure successful calibration within a single scenario. Here, calibration is conducted with all but one sensor's data, which is then used to assess the calibration's accuracy.

\subsection{Main Experiment: leave-one-scenario-out} \label{sec:main_exp}
The leave-one-scenario-out experiment consists of two stages. In the training stage, the model is calibrated using a set of scenarios and corresponding reference pressure head measurements. In the test stage, the change in simulation versus measurement error is calculated for a scenario not included in the training.

This is the primary experiment because it enables a reliable comparison of the calibration effects using two different data sets (e.g., hydrant trials vs. daily usage) on the same test scenario. Additionally, it simulates the real-life process of building and deploying a hydraulic model: first, the model is calibrated and validated with available data, and then it is applied to new, unseen data. 

\subsubsection {Training stage}
In the training stage, the input data includes initial roughness vector $\vect{r}^0$, 
a family of scenarios $\{s_i\}_{i=1}^{k-1}$,
a family of reference pressure head vectors $\{\vect{p}^M_i\}_{i=1}^{k-1}$, a subset of nodes used for calibration algorithm $J=M$, a subset of nodes used for optimization stage of calibration $O \subseteq N$ ($O = N$ for ANN-PSO method, $O = J$ for clustering-COBYLA method).

Training involves determining the roughness vector for the graph $\mathcal{G}$, utilizing calibration methods described in Sec.~\ref{sec:calibration_procedure}. The result of this stage is calibrated pipe roughness coefficients vector~$\vect{r}$.

\subsubsection {Test stage} \label{sec:test_stage}
In the test stage, the input data includes
test scenario $s_k$, reference pressure head vector $\vect{p}^M_k$, initial roughness coefficients vector $\vect{r}^0$, roughness coefficients vector from the training stage $\vect{r}$, subset $J'=M$ of nodes used for the test stage.

The test stage procedure is as follows:
\begin{enumerate}   
\item     Run the simulation of non-calibrated hydraulic model; the input to the simulation $\mathbb{S}_{\mathcal{G}}$ is the test scenario $s_k$ and the initial roughness vector $\vect{r}^0$; obtain the simulated pressure head vector $\vect{p}^0\in \R^n$ from non-calibrated hydraulic model;\\
\item Run the simulation of calibrated hydraulic model; the input to the simulation $\mathbb{S}_{\mathcal{G}}$ is the test scenario $s_k$ and the roughness vector $\vect{r}$ resulting from the training stage; obtain the simulated pressure head vector $\vect{p}\in \R^n$ from calibrated hydraulic model;\\
\item Extract the subset vector $\vect{p}^0_{J'}$, representing $J'$ subset of nodes, from the vector $\vect{p}^{0}$ , and analogically extract the vector $\vect{p}_{J'}$ from the vector $\vect{p}$; \\
\item Calculate the absolute error vector $\vect{e}^0$ of non-calibrated hydraulic model:
    \begin{equation}
        \vect{e}^0=|\vect{p}^0_{J'}-\vect{p}^{M}_{J'}|
    \end{equation} 
\item Calculate the absolute error vector $\vect{e}$ of calibrated hydraulic model:
    \begin{equation}
        \vect{e}=|\vect{p}_{J'}-\vect{p}^{M}_{J'}|
    \end{equation}
\item Calculate the change (improvement) of absolute error $\vect{\Delta e}$ compared to the state before calibration:
    \begin{equation}
         \vect{\Delta e}=\vect{e}^0-\vect{e}
        \label{eq:delta_e}
    \end{equation}    
	\end{enumerate}
The result of this stage is the change of error $\vect{\Delta e}$, which measures the amount of improvement of a model fit achieved by a given calibration procedure.

\subsubsection{Experimental setup} \label{subsec:exp_setup}

The experiments aim to confirm that calibration using hydrant trial data provides greater uncertainty reduction compared to standard calibration based on daily usage data. To achieve this, we conduct the leave-one-scenario-out experiment in various configurations distinguished by different input data sets and calibration methods. The abbreviations are as follows: 'H' refers to the Hydrant Trials set, 'D' to the Daily Usages set, 'AP' to the ANN-PSO algorithm, and 'C' to the clustering-COBYLA algorithm.

Each experimental setup is named using three abbreviations: the first indicates the training data set, the second the test data set, and the third—preceded by a dash—indicates the calibration method used. For example, $HH-AP$ represents a fourfold experiment where calibration is trained and tested on hydrant trials using the ANN-PSO method.

In relation to the central thesis of the article, HH and DH experimental setups are the most critical. In the HH setup, training is performed on three hydrant trial scenarios, and the test is performed on the fourth, left out, hydrant trial scenario. This group of experiments represents our proposed approach to improve calibration with the use of short-lasting hydrant trials. The DH setup involves training on four daily usage scenarios and the test on the single hydrant trial scenario. It is treated as a reference to the HH group because it represents a classical approach to calibration on normal operating conditions. Since the calibration results are evaluated within the same hydrant trial test scenarios, it is permissible to compare the test results between the HH and DH experiments, thereby enabling the verification of the article's central thesis.

The experiments use the following parameter values:
\begin{enumerate}
    \item ANN: layer structure: 16-32-100, max. number of epochs: 500, batch size: 32, learning rate: $10^{-3}$, solver: SGD solver, training set size: 100. While building the training set, the roughness range is $(0.01,10)$ 
    \item PSO: $w =  0.1$,  $c_1 = 1.0$,  $c_2 = 1.0$ (based on \cite{vaz2013benchmark}),  maximal no. of iterations: 100, no. of particles: 32,   Bratton-Kennedy bounding method \cite{bratton2007defining},  roughness bounds: $(0.01,10)$
    \item Clustering-COBYLA (parameters selected following \cite{powell1994direct}): pipes roughness and flow rate as KMeans clustering criterion, roughness bounds: $(0.01,10)$ and stopping criterion: error of $10^{-4}$, maximal no. of iterations: 300.
\end{enumerate}

\subsubsection{Measures for experiments} \label{subsec:meas_exp}    
The assessment of the impact of hydrant tests versus daily consumption is quantified by the parameter $\vect{z}_{i}$,
\begin{equation}
    \vect{z}_{i}= \vect{\Delta e}^{DH(i)}- \vect{\Delta e}^{HH(i)}, i=\{1,...,k\},
\end{equation}
where  $\Delta \vect{e}^{DH(i)}$ is the change of error in DH experiment with the use of $i$-th hydrant trial test scenario, $\vect{\Delta e}^{HH(i)}$ is the change of error in HH experiment with the use of $i$-th hydrant trial test scenario and $\vect{z}_{i}$ is length $m$ coefficients vector that indicates the extent of improvement in the approach utilizing hydrant trials compared to the baseline approach based on daily consumption. While using the definition of $\vect{\Delta e}$ (as in Eq.~\ref{eq:delta_e}), the $\vect{z}_{i}$ vector  can be further represented as:
\begin{equation}
    \vect{z}_{i}= \vect{e}^{0(i)}-\vect{e}^{DH(i)}- (\vect{e}^{0(i)}-\vect{e}^{HH(i)})=\vect{e}^{HH(i)}-\vect{e}^{DH(i)}
    \label{eq:z_i}
\end{equation}
where $\vect{e}^{0(i)},\vect{e}^{DH(i)},\vect{e}^{HH(i)}$ denotes error in the $i$-th test hydrant trial scenario between simulated and measured pressure head, with abbreviations as follows: 0 - before calibration, DH - after calibration with daily usage scenarios, HH - after calibration with hydrant trial scenarios. The final formula from Eq.~\ref{eq:z_i} allows for omitting the error before calibration in the calculation of $\vect{z}_{i}$ for direct comparison of errors after calibration using DH and HH experiment approaches.

\subsection{Supplementary Experiment} \label{sec:suppl_exp}
In this experiment, referred to as leave-one-sensor-out, a single scenario with corresponding reference pressure heads from measurements is used as a total input data set -- all but one sensor readings are used for the calibration, and the left-out sensor is used for measuring the change of error due to calibration.

\subsubsection{Training stage}		
 Input data to the training stage include initial roughness vector $\vect{r}^0$,
one-element training scenario family $\{s\}$,
one-element reference pressure head family $\{\vect{p}^M\}$, set $J = M \setminus\{x\}$ where $x\in M$ is the sensor node 'left out' for the test stage, parameter $O \subseteq N$ ($O = N$ for ANN-PSO method, $O = J$ for clustering-COBYLA method). The training stage involves determining the roughness vector~$\vect{r}$ for $\mathcal{G}$, utilizing calibration methods described in Sec.~\ref{sec:calibration_procedure}.
\subsubsection{Test stage}
The test stage is realised with the following input: test scenario $s=(p^R, \vect{d})$ (the same one as in the one-element family in the training stage), reference pressure head vector $\vect{p}^M$ (the same one as in the one-element family in the training stage), one-element set $J' = \{x\}$ containing the test sensor node.
The test stage procedure is realised analogically as described in Sec.~\ref{sec:test_stage}. The result of this stage is the change of error calculated from Eq.~\ref{eq:delta_e}.

\section{Results} \label{sec:results}
This section presents the results of the main leave-one-scenario-out experiment.This machine learning framework enables a clear comparison of the impact of hydrant trials versus daily consumption scenarios on calibration effectiveness by evaluating the same test dataset.

From Eq.~\ref{eq:z_i} it can be concluded that the positive values of elements of $\vect{z}_{i}=\begin{bmatrix} z_{ij} \end{bmatrix}$, corresponding to the change of error on sensor nodes in the test scenario, mean that calibration on hydrant trials gave greater reduction of error when comparing to baseline calibration on daily usage. The aggregate $\bar{z}=\frac{1}{km}\sum_{i=1}^{k}\sum_{j=1}^m z_{ij}$ is the average gain when using the proposed approach. For the `clustering-COBYLA', $\bar{z}=(0.001 \pm 0.002)\:\mathrm{m}$, for `ANN-PSO' $\bar{z}=(0.0009 \pm 0.003)\:\mathrm{m}$, both showing consistent improvement. Statistical testing of individual $z_{ij}$ values with the Wilcoxon signed-rank test (used instead of t-test, as the Shapiro-Wilk test of $z_{ij}$ rejects the null hypothesis with p-value of $0.00002$ for `clustering-COBYLA' and $0.02$ for `ANN-PSO') provides an additional verification, confirming statistical significance for both `clustering-COBYLA' (p-value $0.00004$) and `ANN-PSO' (p-value $0.05$).

\begin{figure}[p]
\begin{subfigure}{1\linewidth}
\centering
\includegraphics[width=0.71\textwidth]{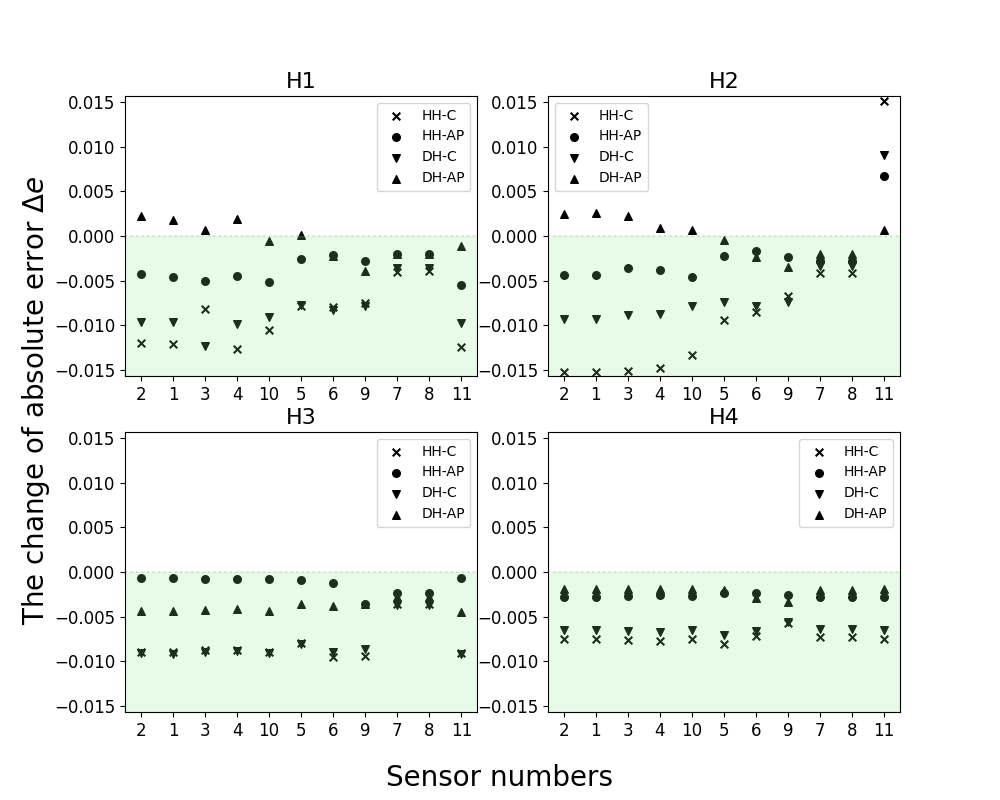}
\caption{The results of HH and DH experiments on test hydrant trials H1, \dots, H4.}
\label{fig:delta_e_lo_scenario_out_hp}
\end{subfigure}
\begin{subfigure}{1\linewidth}
\centering
\includegraphics[width=0.71\textwidth]{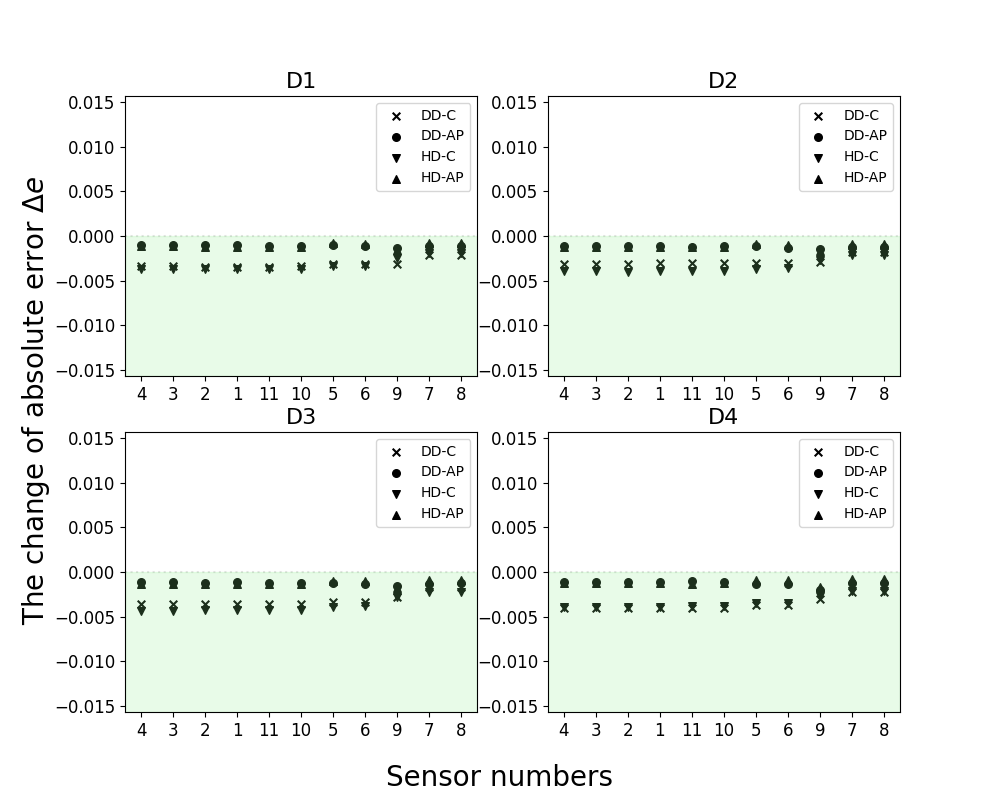}
\caption{The results of HD and DD experiments on daily usage test scenarios D1, \dots, D4.}
\label{fig:delta_e_lo_scenario_out_ref}
\end{subfigure}
\caption[The results of the main leave-one-scenario-out experiments]{The change of the absolute error $\vect{\Delta e}$ in leave-one-scenario-out experiments. The abbreviations are as follows: 'HH'--experiment with hydrant trials training set and unseen hydrant trial test set; 'DH' --experiment with daily usages training set and hydrant trial test set; 'HD'-- experiment with hydrant trials training set and daily usage test set; 'DD' --experiment with daily usages training set and unseen daily usage test set; 'AP'--ANN-PSO algorithm; 'C'--clustering-COBYLA algorithm (refer to Sec.~\ref{subsec:exp_setup}). Green-shaded area marks a region of negative $\vect{\Delta e}$, which is also the area of effective error reduction due to calibration. Sensor numbers in (a) and (b) are sorted by experimental scatter series of the biggest error reduction}
\end{figure}

The detailed results for individual sensors and test scenarios in HH and DH experiments lead to several conclusions, as shown in Fig.~\ref{fig:delta_e_lo_scenario_out_hp}. The Figure consists of four subplots representing four hydrant trial test scenarios, labelled 'H1', ..., 'H4'. Each sub-figure presents the change of absolute error $\vect{\Delta e}_i=[\Delta e_{ij}]=f(i)$, where  $\Delta e_{ij}$ denotes the change of absolute error on i-th test scenario and j-th sensor and corresponds to a single point of the subplot. The X-axis shows the numerical designations of the sensors from the set \{1,2,..,11\}. Sensor numbers are globally sorted by the HH-C series on the 'H2' subplot, which includes the lowest negative $\Delta e_{ij}$ value, to improve visualization clarity. The green-shaded area on the subplots indicates the region of negative $\Delta e_{ij}$ values. The negative $\Delta e_{ij}$ value signifies that calibration successfully reduced the absolute error (between simulation and measurement pressure head) in the j-th sensor of the i-th test scenario compared to the initial state. 

From Fig.~\ref{fig:delta_e_lo_scenario_out_hp}, it can be concluded that both calibration approaches ('AP' -- ANN-PSO and 'C' -- clustering-COBYLA) lead to error reduction on most sensors of the test scenarios -- points are located within the green-shaded area. As an extension of the aggregated results presented earlier in this Section, the Figure shows that calibration using hydrant trials (HH experiments) generally leads to more significant error reduction in the test scenarios compared to calibration based on daily usage scenarios. (DH experiments), regardless of which calibration method ('AP' or 'C') is used. This can be observed as the 'HH' point of the j-th sensor is located lower than the 'DH' point of the same sensor.  

The most significant error reductions due to hydrant trial-based calibration occur in the 'far' test scenarios ('H1' and 'H2', refer to Sec.~\ref{sec:dataset}), with the best result being $-0.015\:\mathrm{m}$ on sensor 2 in the 'H2' scenario. This corresponds to a 45\% reduction from the initial error of $0.033\:\mathrm{m}$.  

These scenarios also show the most significant benefit of hydrant trial calibration over daily usage calibration, with differences between corresponding $\Delta e_{ij}$ from 'DH' and 'HH' points exceeding 0.005~m on several sensors. This is because hydrant trials located farther from the inflow source increase flow velocity and pressure gradient over a larger area of the WDN.

For the 'close' scenarios ('H3' and 'H4'), the advantage of 'HH' over 'DH' is less pronounced. In the 'H3' scenario, the 'AP' case shows greater error reduction for 'DH' points, while in the 'C' case, 'HH' points show slightly better performance. In the 'H4' scenario, 'HH' points are consistently lower than 'DH', but the overall benefit from using hydrant trials is minimal, with differences in $\Delta e_{ij}$ below $ 0.001\:\mathrm{m}$.

It is important to note that when sorting the plots by the series with the most significant error reduction, a trend emerges in the 'H1' and 'H2' subplots. The 'left' side shows a clearer predominance of 'HH' results over 'DH', with more profound error reduction, while the 'right' side does not always follow this trend, and in some cases, the opposite occurs. This indicates that specific sensors (e.g., sensors 2, 1, 3, 4) respond better to calibration. However, this holds only for the 'far' scenarios.

In terms of the calibration approaches used, Fig.\ref{fig:delta_e_lo_scenario_out_hp} shows that the 'C' method achieves significant error reduction through hydrant trials and, in some cases, through daily usage scenarios (on several sensors in the 'H1', 'H2', and 'H3' scenarios, 'DH-C' points reach $-0.01\:\mathrm{m}$). This method consistently highlights the effectiveness of 'HH' experiments over 'DH' ones, with only a few sensor-specific exceptions. The 'AP' calibration, when using hydrant trials, results in error reduction of up to $-0.005\:\mathrm{m}$ in the 'H1' and 'H2' test scenarios. However, when daily usage data is utilized for calibration, some sensor points in the 'H1' and 'H2' cases show increased error.

Some 'problematic' sensors can be identified, such as sensors 7 and 8 in the 'H1', 'H2', and 'H3' test cases. These sensors show little error reduction regardless of the calibration data set or method. In the 'H2' test scenario, sensor 11 consistently experiences an undesirable increase in error following calibration.

While calibration effects are most pronounced in scenarios with increased flow and pressure gradients, such as hydrant trials, we also examine the calibration effects on daily usage test scenarios, as hydraulic models are routinely used to simulate normal operating conditions. These details are discussed in Sec~\ref{sec:discussion}. 

To further assess calibration efficiency, a supplementary leave-one-sensor-out experiment was conducted within individual hydrant trials or daily usage scenarios. The results are also presented in Sec.~\ref{sec:discussion}.

\section{Discussion} \label{sec:discussion}
\subsection{Results in the context of existing literature}
In our study, the average error across all hydrant trial scenarios before calibration is $0.102\: \mathrm{m}$, and in the best case (HH-C experiment), this is reduced to $0.094\: \mathrm{m}$. This may seem modest relative to typical calibration errors. For example, \cite{shiu2024enhancing} reports pressure errors in a calibrated network exceeding $0.1 \: \mathrm{kg/cm^2}$ (approximately $1 \: \mathrm{m H_2O}$), even with simulated sensor coverage of 30\% of the network. In \cite{lynnfield2024}, a difference of around $3.5\:\mathrm{m}$ between simulated and measured pressure values is considered highly accurate. Similarly, \cite{zanfei2020calibr} reports a mean absolute error (MAE) of $0.28\:\mathrm{m}$ in a small network, which is deemed promising. However, these results are from networks with high-pressure gradients, where such calibration precision is sufficient for leak detection, a primary goal in hydraulic model calibration.  In the literature, many works focus on high gradient WDNs \cite{vrachimis2022battledim}. 

In contrast, in our study area, as shown in Fig.\ref{fig: hydrant_trials}, even a significant outflow from hydrant trials (10$\mathrm{m^3/h}$) results in less than $1 \: \mathrm{mH_2O}$ of pressure head drop at sensor locations. A comparable network in terms of pressure head values, though smaller in scale, is the Modena network studied in \cite{zanfei2020calibration}, where the authors report a mean absolute error of 0.9~m after calibration, which is ten times higher than our result.

\subsection{Practical utilization of the results}
Our short-duration hydrant trial calibration approach enables significant water savings compared to traditional methods. For example, \cite{sophocleous2017two} describes four parallel hydrant trials with total discharges ranging from $8.27\:\mathrm{l/s}$ to $9.89\:\mathrm{l/s}$, lasting 4.5 hours. Assuming an average flow rate of $9.08\:\mathrm{l/s}$, equivalent to $32.69\:\mathrm{m^3/h}$, the estimated water loss during that survey is $147\:\mathrm{m^3}$.

In contrast, our hydrant trials, lasting approximately 15 minutes each (including discharge stabilization), resulted in an estimated total water loss of $10\:\mathrm{m^3}$. This represents a savings of $137\:\mathrm{m^3}$. Given water prices ranging from €0.4 to €1.2 per $\mathrm{m}^3$ \cite{vrachimis2022battledim}, the financial savings range from €55 to €164.

\subsection{Leave-one-scenario-out with daily usage test scenarios} \label{sec:lo_scenario_day}
In contrast to hydrant trial scenarios, during normal operating conditions, the study area exhibits low-pressure gradients. This indicates minimal calibration effects, as even significant variations in roughness produce minimal head loss changes. This is confirmed through the leave-one-scenario-out experiments involving daily usage test scenarios ('HD' and 'DD' setups, see Sec.~\ref{subsec:exp_setup}), as illustrated in Fig. \ref{fig:delta_e_lo_scenario_out_ref}. Calibration on hydrant trial scenarios ('HD') did not lead to a greater reduction in absolute error compared to calibration on daily usage scenarios ('DD'), as evidenced by the nearly overlapping 'HD' and 'DD' points for the same calibration method ('C' or 'AP').

\subsection{Supplementary experimental results}
\begin{figure}[p]
\begin{subfigure}{1\linewidth}
\centering
\includegraphics[width=0.71\textwidth]{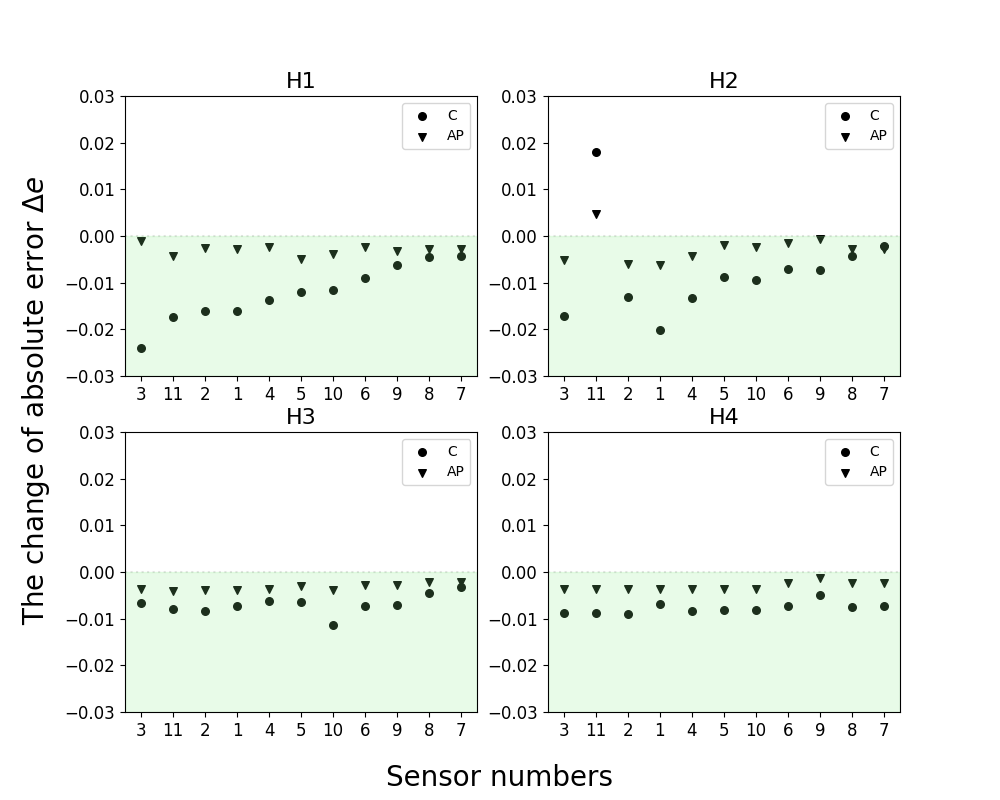}
\caption{Hydrant Trials}
\label{fig:delta_e_lo_sensor_out_hp}    
\end{subfigure}
\begin{subfigure}{1\linewidth}
\centering
\includegraphics[width=0.71\textwidth]{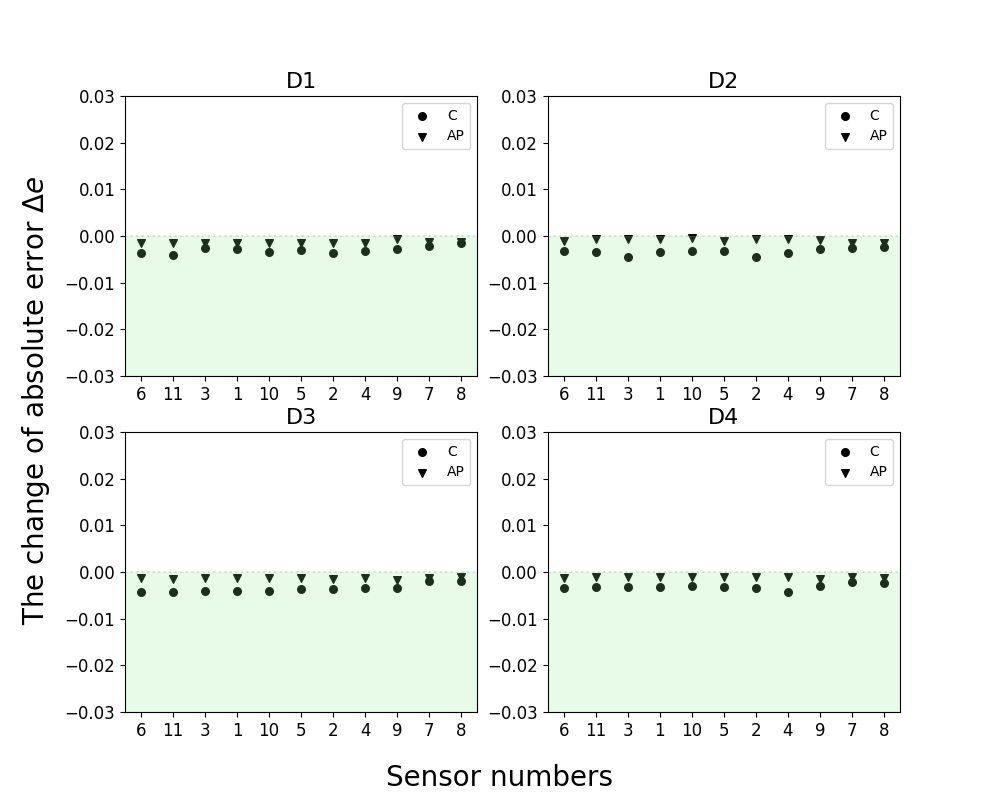}
\caption{Day Reference}
\label{fig:delta_e_lo_sensor_out_ref}    
\end{subfigure}
\caption[The change of the absolute error in leave-one-sensor-out experiments]{The change of the absolute error $\vect{\Delta e}$ in leave-one-sensor-out experiments on test sensor in (a) hydrant trials scenarios H1, \dots, H4; (b) daily usage scenarios D1, \dots, D4. Each point represents the test stage result of the experiment with the test ('left') sensor number indicated on the X-axis. 'AP' stands for ANN-PSO algorithm, while 'C' stands for clustering-COBYLA algorithm. Green-shaded area marks a region of negative $\vect{\Delta e}$, which is also the area of effective error reduction due to calibration. Sensor numbers are sorted by experimental scatter series of the biggest error reduction.}
\label{fig:mae_lo_sensor_out}
\end{figure}

The supplementary leave-one-sensor-out experiment evaluates the impact of calibration, performed using a subset of sensor nodes, on an unseen sensor node within a hydraulic graph for a single scenario.
While this experiment does not allow a direct comparison between hydrant trial-based and daily usage-based calibration (as the test sets differ), it provides insight into the calibration effects within the same scenario and time period. Fig.~\ref{fig:delta_e_lo_sensor_out_hp} and Fig.~\ref{fig:delta_e_lo_sensor_out_ref} show the change in absolute error for leave-one-sensor-out experiments on hydrant trial and daily usage scenarios, respectively. Each point represents the result for a single sensor, with the X-axis indicating the sensor used in the test set.
From Fig.~\ref{fig:delta_e_lo_sensor_out_hp} we can conclude that calibration using a 'far' hydrant trial ('H1' and 'H2') reduces the error by up to $-0.025\: \mathrm{m}$ (in the 'H1' scenario), while 'close' hydrant trials ('H3' and 'H4') yield error reductions slightly below $-0.01\: \mathrm{m}$ (in the 'H3' scenario). In contrast, Fig.\ref{fig:delta_e_lo_sensor_out_ref} shows that daily usage scenarios do not achieve similarly notable reductions (error changes do not fall below $-0.05\: \mathrm{m}$).
These observations align with the main results (Sec.~\ref{sec:results}) and the leave-one-scenario-out tested on daily usage scenarios (Sec.~\ref{sec:lo_scenario_day}), highlighting the improved error reduction achieved within hydrant trial scenarios compared to daily usage scenarios.

\subsection{ANN regression in ANN-PSO method}
\begin{figure}[ht]
\centering
\includegraphics[width=0.8\linewidth]{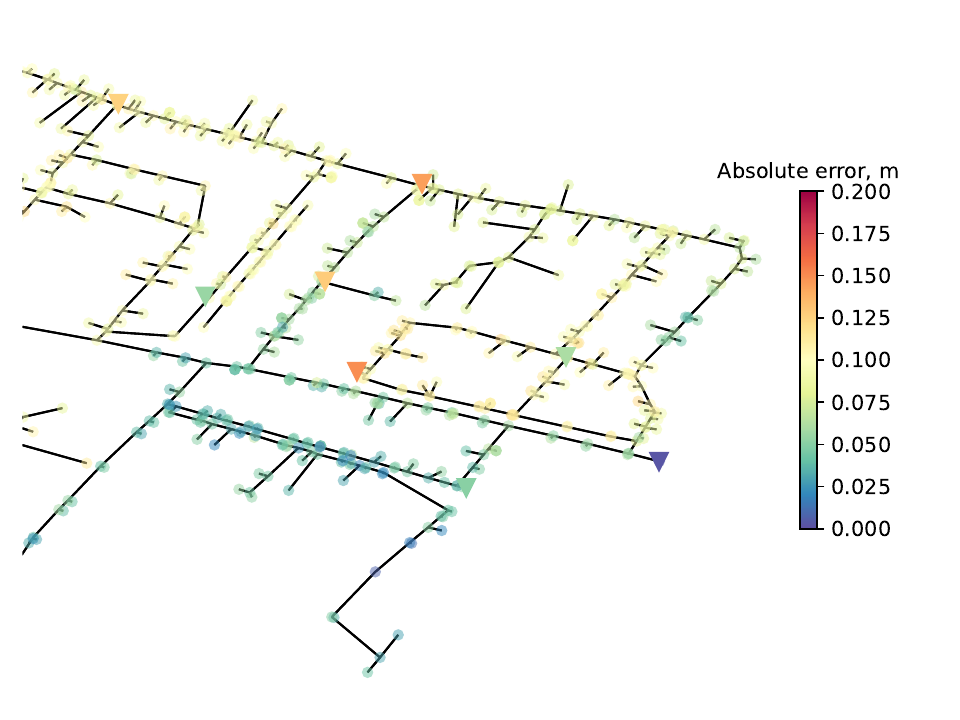}
\caption[Visualization of the study area graph fragment with results of ANN-PSO calibration approach on example scenario from the training stage of the main experiment]{Visualization of the study area graph fragment with results of ANN-PSO calibration approach on example scenario from the training stage of the main experiment. Nodes marked as $\circ$ represent the absolute error between the pressure head simulated from the calibrated hydraulic model and the pressure head estimated by ANN. Nodes marked as $\triangledown$ represent the absolute error between the pressure head simulated from the calibrated hydraulic model and the pressure head measured in the sensor points.}
\label{fig:ann.png}
\end{figure}

An interesting observation arises from the ANN-PSO approach. We compared the results of the ANN regression stage, where the pressure head vector (combining measured and regressed values) is used, with the simulation results after ANN-PSO calibration. A fragment of the graph showing absolute error values per node is presented in Fig.~\ref{fig:ann.png}. 

Nodes marked as $\circ$ indicate the absolute error at ANN estimation points, while nodes marked as $\triangledown$ represent the error at sensor points. Certain sensor nodes exhibit increased error compared to nearby regions, while others show lower error than surrounding nodes. This suggests that the ANN regression may not have adapted well to measurements in specific areas. A limitation here stems from hydraulic equations—ANN training is based on hydraulic simulations. If measurements, subject to error, deviate from the hydraulic gradient, the ANN may struggle to accurately reconstruct the gradient, leading to local errors.

This issue may also indicate the need to improve the MLP component, aligning with the statement from \cite{meirelles2017} that ANN performance is crucial to the overall method. Since ANN-estimated nodes far outnumber measured nodes, the optimization process may overly favour ANN approximations, potentially reducing the influence of ground-truth data.  

\subsection{Alternative approach to utilizing training scenarios in ANN-PSO method}
Since \cite{meirelles2017} does not specify a detailed method for handling multiple scenarios in the ANN-PSO approach, we perform the ANN regression stage separately for each scenario in the training set, as outlined in Sec.~\ref{sec:calibration_procedure}. We tested two methods for passing the ANN regression results from multiple scenarios to the PSO stage, named 'Before' and 'After.'

In the 'Before' method, a set of pressure head vectors $\{\vect{p}_{O,i}\}_{i=1}^k$ from ANN regression stage creates a single input into a PSO run. This means PSO optimizes the simulation based on multiple input vectors, resulting in a single roughness vector $\vect{r}$.

In the 'After' method, PSO is run separately for each pressure head vector $\vect{p}_{O,i}$, derived from ANN regression. Each PSO process generates an individual roughness vector $\vect{r_i}$. This set of roughness vectors $\{\vect{r}_i\}_{i=1}^k$ is used to calculate the mean absolute error (MAE) between simulated and measured pressure heads for each train scenario $s_i$. The roughness vector $\vect{r}$ that yields the lowest average MAE is selected as the final result. 

The results from both methods are similar, regardless of the training scenario set used. In the Sec.~\ref{sec:results}, we use the "Before" setting.

\subsection{Parameters of PSO stage in ANN-PSO approach} \label{sec:PSO_params}
\begin{figure}[ht]
\begin{subfigure}[t]{0.45\linewidth}
\centering
\includegraphics[width=1\textwidth]{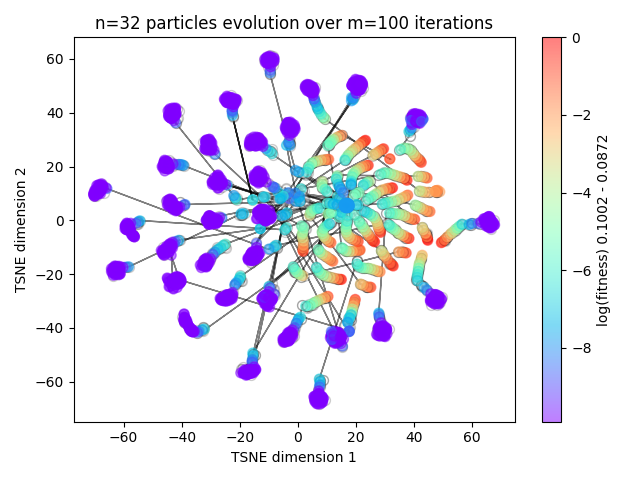}
\subcaption{Correct PSO process for the set of parameters chosen for experiments: $c_1=1.0$, $c_2=1.0$, $w=0.1$}
\label{fig:pso_correct_chosen}    
\end{subfigure} \hfill
\begin{subfigure}[t]{0.45\linewidth}
\centering
\includegraphics[width=1\textwidth]{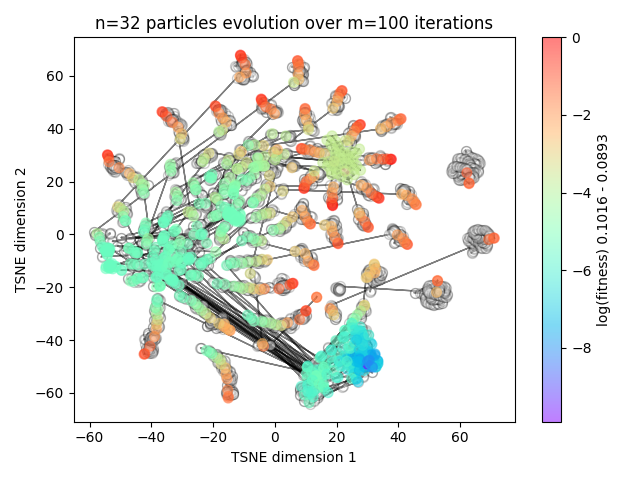}
\subcaption{Correct PSO process for the set of parameters: $c_1=0.7$, $c_2=0.7$, $w=0.5$}
\label{fig:pso_correct_1}    
\end{subfigure}\\
\begin{subfigure}[t]{0.45\linewidth}
\centering
\includegraphics[width=1\textwidth]{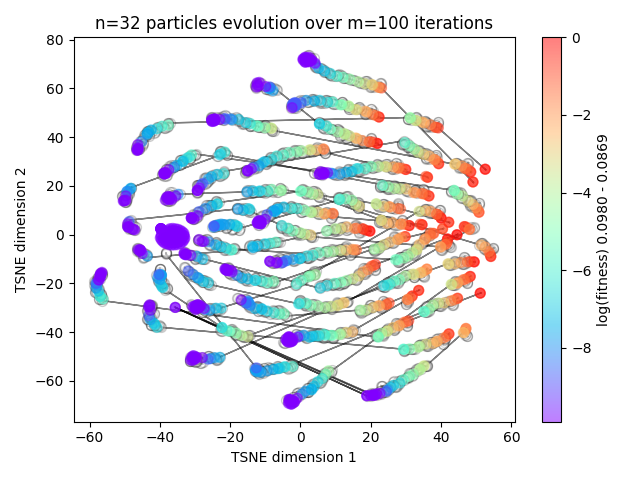}
\subcaption{Correct PSO process for the set of parameters: $c_1=0.5$, $c_2=0.5$, $w=0.5$}
\label{fig:pso_correct_2}    
\end{subfigure} \hfill
\begin{subfigure}[t]{0.45\linewidth}
\centering
\includegraphics[width=1\textwidth]{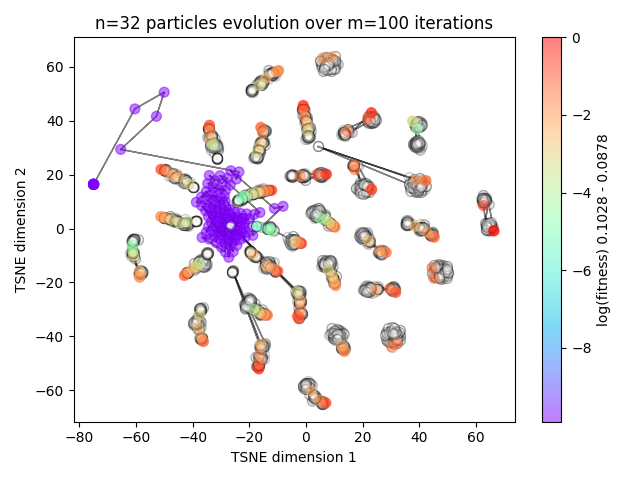}
\subcaption{Incorrect PSO process for the set of parameters: $c_1=1.0$, $c_2=1.0$, $w=0.3$}
\label{fig:pso_incorrect}    
\end{subfigure}
\caption[Examples of PSO swarm movement visualizations]{Examples of correct (a, b, c) and incorrect (d) PSO swarm movement visualizations based on parameters from \cite{vaz2013benchmark}}
\label{fig:pso_history}
\end{figure}

We initially tested the ANN-PSO algorithm using the parameters provided by \cite{meirelles2017}, but the results on our dataset, due to its specific characteristics, were unsatisfactory. To improve performance, we conducted a literature-based grid search. After analyzing swarm movement during calibration in a specific scenario (see Fig.~\ref{fig:pso_history}), we selected parameter sets from existing literature that exhibited the desired swarm behaviour and convergence \cite{vaz2013benchmark}: $c_1=1.0$, $c_2=1.0$, $w=0.1$; $c_1=0.7$, $c_2=0.7$, $w=0.5$; and $c_1=0.5$, $c_2=0.5$, $w=0.5$. Ultimately, we chose the set $c_1=1.0$, $c_2=1.0$, $w=0.1$ due to its lowest Mean Absolute Error.

\section{Conclusions}
This article presents a methodology for calibrating WDN hydraulic models using brief nighttime hydrant trials. We developed two machine learning-inspired cross-validation frameworks to evaluate calibration efficiency on a unique case study dataset. Our results demonstrate that this method improves uncertainty reduction compared to traditional daily usage-based calibration, regardless of the calibration algorithm used.

The novelty of our approach lies in the short duration of increased water discharge, which minimises water loss during hydrant trials. This method has the potential to appeal to water utilities and management practitioners, offering improved hydraulic model calibration with minimal financial risk compared to long-term hydrant discharges.

\section{Data Availability Statement}
Some data, models, or codes that support the ﬁndings of this study are available from the corresponding author upon reasonable request.

\section{Acknowledgment}
This work has been partially supported by the Polish National Centre for Research and Development grant POIR.01.01.01-00-1414/20-00, "Intelligence Augumentation Ecosystem for analysts of water distribution networks".

\bibliographystyle{plain}
\bibliography{bibliography}
\end{document}